\documentclass[12pt,reqno]{amsart}
\allowdisplaybreaks[4]
\usepackage{graphicx}
\usepackage{lipsum}
\usepackage{amssymb}
\usepackage{color}
\usepackage{amsmath}
\usepackage{cite}
\usepackage{subfigure}
\usepackage{graphicx}
\usepackage{epstopdf}
\usepackage{bibentry}
\usepackage{cases}

\newtheorem{remark}{Remark}
\newtheorem{thm}{Theorem}

\begin{document}
\title{Asymptotic dynamics of higher-order lumps in the Davey-Stewartson II equation}
\author{Lijuan Guo$^{1}$, P.G. Kevrekidis$^{2}$, Jingsong He$^{3,\ast}$}
\thanks{$^*$ Corresponding author: hejingsong@szu.edu.cn, jshe@ustc.edu.cn}
\dedicatory {$^{1}$College of Science, Nanjing Forestry University, Nanjing, Jiangsu, 210037, P.\ R.\ China\\
$^{2}$Department of Mathematics and Statistics, University of Massachusetts, Amherst, MA 01003-4515, USA
$^{3}$Institute for Advanced Study, Shenzhen University, Shenzhen, Guangdong, 518060, P.\ R.\ China\\
}

\begin{abstract}
A family of higher-order rational lumps on non-zero constant background of  Davey-Stewartson (DS) II equation are investigated.
These solutions have multiple peaks whose heights and trajectories are approximately given by asymptotical analysis.
It is found that the heights are time-dependent and for large time they approach the same constant height value of
the first-order fundamental lump.
The resulting trajectories are considered and it is found that the
scattering angle
can assume arbitrary values in the interval of $(\frac{\pi}{2}, \pi)$
 which is markedly distinct from the necessary orthogonal scattering
for the higher-order lumps on zero background.  Additionally, it is
illustrated that  the higher-order lumps containing multi-peaked
$n$-lumps
can be regarded as a nonlinear superposition of $n$ first-order ones as $|t|\rightarrow\infty$.

\end{abstract}
\maketitle
\noindent{{\bf Keywords}: Davey-Stewartson~II equation, Darboux transformation, Lump, Asymptotic analysis.}$\\$

\section{Introduction}
In this paper we consider the Davey-Stewartson (DS) II system,
which was first derived by A. Davey and K. Stewartson to model water
waves with weak surface tension~\cite{Davey}. This can also be considered
as a long wave limit of Benney-Roskes equation~\cite{Benney} of the form:
\begin{equation}\label{DSII}
\begin{aligned}
&iu_t+u_{xx}-u_{yy}+(2\kappa|u|^2+S)u=0,\\
&S_{xx}+S_{yy}=-4\kappa(|u|^2)_{xx}, \quad \kappa=\pm 1,
\end{aligned}
\end{equation}
where $u$ is the amplitude of a surface wave packet and $S$ characterizes the mean
motion generated by this surface wave.
A recent discussion of the derivation of such models and their
multiscale expansion connections can be found in the book
of~\cite{mark2011}.
Apart from the realm of water waves, relevant models
can be found to be relevant in other physical fields,
such as nonlinear optics~\cite{Newell,PRE2001,EPJ2007}, plasma physics~\cite{Panguetna,Musher,Nishinari} and
ferromagnets~\cite{Leblond}.
The system is integrable
in that it admits Lax pair (see Eqs.(6.1.2-6.1.3) in Ref.~\cite{matveev}) and can
be solved via inverse spectral transformation with the help of the so-called $\bar{\partial}$
methods~\cite{JMP25}. With regard to the solutions to DS II Eq.~(\ref{DSII}), the defocusing case ($\kappa=1$) only admits line solitons, but
does not possess lump solutions, as proved in Ref.~\cite{Beals}. Consequently,
we limit our attention to the focsuing case ($\kappa=-1$) to derive
higher-order lumps and analyze their dynamics.

Lumps, as a class of rational soltion solutions, are localized in the all
space and
travel in time.
An interesting topic in the realm of soliton dynamics (especially, in
connection
to such higher-dimensional settings)
is to look at the scattering properties of two or more lump solitons colliding.
The simplest type of interaction lumps was first
  discovered by Manakov {\it et al.} in the Kadomtsev-Petviashvili (KP) equation
by employing the dressing method~\cite{PLA1977}. Subsequently
Satsuma and Ablowitz~\cite{JMP20}
 used direct  and long-wave limit methods to construct classes of
 lumps of the KP and DS equations.
 These solutions feature a trivial interaction, i.e., they consist of $n$ lumps traveling with
  distinct asymptotic velocities and their trajectories remain unchanged
before and after interaction (i.e., for large time).  In other words, they  experience
a normal scattering \cite{Stepanyants3} and correspond to $n$-simple pole cases. Such interaction lump solutions on zero-boundary background
of DS II were also obtained
by Arkadiev, Pogrebkov and Polivanov via
the inverse scattering method~\cite{Physd1989}.
However, if the individual lumps have the  same asymptotic center-of-mass velocities, they  undergo anomalous scattering (an infinite
 phase shift of their trajectories)
 with a non-zero deflection angle after a head-on collision. These
 correspond to higher-order poles \cite{Villarroel3,Villarroel4,PLA2000}.

 Many authors have also used different methods to study higher-order lumps of KP-I previously~\cite{Stepanyants3,Nonlinearity26,chenshihua2016,Chang2018,Yang2022,Chakravarty2022,ling2022,
Stepanyants1,KPIarxiv}. Gorshkov {\it et al.}~\cite{Stepanyants3} reported a second-order lump
solution which describes the nontrivial interaction and anomalous scattering of two lumps, which defied the paradigm of solitons as non-interacting entities. Ablowitz {\it et al.}\cite{PLA2000}
used the inverse scattering transformation and binary Darboux Transformation (BDT) to construct higher-order lumps that
include the solution of \cite{Stepanyants3} as a special case. They
found that when $t$ runs from $-\infty$ to $0$, these $n$ lump peaks
first attract each other and overlap, after which time they experience a large angle scattering, then again separate into $n$ peaks as $t\rightarrow+\infty$.
Other integrable equations such as the Boussinesq
equation~\cite{physd1995,Clarkson2017},  the $2+1$-dimensional NLS equation~\cite{Villarroel1,Villarroel2},
$2+1$-dimensional asymmetric Nizhnik-Novikov-Veselov
system~\cite{GuoJPA2021} and the $2+1$-dimensional chiral
equation~\cite{Ward,Ioannidou}  
have also been found to feature similar
solution structures. In
Refs.~\cite{Chang2018,Yang2022,Chakravarty2022,ling2022},
the authors further found that
the higher-order lumps split into a certain number of fundamental ones whose relative spatial separation grows in
proportion to $|t|^{q}$ where $\frac{1}{3}\leq q\leq \frac{1}{2}$ as $|t|\rightarrow\infty$.

However, up to now, the asymptotic dynamics and scattering
  phenomena of the higher-order lumps of DS II equation were  studied,
  to our knowledge, on a vanishing background. Ma$\tilde{\rm n}$as and
Santini studied a large class of higher-order lumps on the zero background of the DS II equation with the use of a
Wronskian scheme~\cite{PLA1997} and later different
groups~\cite{SIAM2003,fokasphysd} also used the inverse scattering method
to construct such rational solutions.
They behave highly nontrivially upon interaction (a head-on collision results in a orthogonal scattering).
A natural question arises whether
there exist novel lumps of DS II equation which  feature anomalous scattering phenomena and   
 scatter with non-orthogonal angle after collision. To this end,
we need to construct a family of new rational lump solutions of DS II
on a non-zero background and to explore their interactions which is a
focal point of the present work.

The Darboux transformation (DT) has been used successively to obtain soliton, breather and rogue wave solutions
in the last several years~\cite{xinkou, CNSNS2017,CNSNS2019,Yangjianke1,akhmediev1,akhmediev2,lingliming1,yanzhenya,wanglihong1,wanglihong2,
guoboling1,Degasperis,Mugui}.
Given its earlier success, we utilize this method herein
to construct higher-order rational lump solutions on non-zero constant background
for DS II equation. To realize this goal, first we need to solve the Lax pair equations to find a hierarchy of solutions,
which are used to construct more general DT.
Indeed, one of our key results consists of the confirmation of the
feature that arbitrary order Taylor coefficients of the fundamental
eigenfunction (the usual exponential solution to Lax pair) all satisfy Lax pair equations with the
same plane wave seeding solution.

Motivated by the above results, we shall concentrate on the following results.
\begin{itemize}
  \item Beginning with the plane wave seeding solution, a hierarchy of new eigenfunctions {\color{red}} generated by these Taylor
  coefficients of a usual exponential solution to the Lax pair, which are used to generalize the $n$-fold DT.
  \item Apart from the $\frac{\pi}{2}$ scattering occurring in collision between lumps \cite{PLA1997,SIAM2003},
  we find a family of higher-order lumps on nonzero background of the DS II equation where the scattering angle can
  be an arbitrary constant in the interval of $(\frac{\pi}{2}, \pi)$. The anomalous scattering and the time evolution process are illustrated by analyzing
the approximate asymptotic formula of these lumps' trajectories.
  \item  The approximate heights of these lump peaks evolve
in time and approach the maximum value of the first-order fundamental lump as $|t|\rightarrow\infty$, which
demonstrates how the $n$th-order lumps constitute a superposition of $n$ distinct peaks.
\end{itemize}

The rest of this paper is organized as follows. In Section $2$, we
begin with the plane wave seeding solution, and establish that
the Taylor coefficients of the fundamental eigenfunction all satisfy Lax pair equations.
In Section $3$, the rational lump solutions up to the third-order are obtained by using DT,
and their dynamical properties are studied.  Our conclusions, as well as some potential directions
of future study are given in the final Section.

\section{Eigenfunctions and Darboux transformation}
The DS II Eq.~(\ref{DSII}) admits the following Lax pair equations\cite{matveev}
\begin{equation}\label{lax}
\Psi_y=J\Psi_x+U\Psi,\quad \Psi_t=2J\Psi_{xx}+2U\Psi_x+V\Psi,
\end{equation}
with a constant diagonal matrix $ J=\left(\begin{array}{cc}
{\rm i} &0\\
0 &-{\rm i}
\end{array}\right) $,
and two potential matrices
\begin{equation}
U=\left(\begin{array}{cc}
0 &u\\
v &0
\end{array}\right),\quad
V=\left(\begin{array}{cc}
(w+{\rm i}Q)/2 &u_{x}-{\rm i}u_{y}\\
v_{x}+{\rm i}v_{y} & (w-{\rm i}Q)/2
\end{array}\right).
\end{equation}
Here, the vector $\Psi=(\psi,\phi)^T$ ($T$ denotes transpose),
the potentials $u,~v=- u^*\in \mathbb{C}$, and the field $Q=2\kappa|u|^2+S \in \mathbb{R}$,
are functions of the three independent variables $x, y, t$.

In this work, we restrict our attention to the plane wave seeding solution, i.e.,
\begin{equation}\label{seed}
u=a \exp\big[{\rm i}(bx+cy+dt)\big], \quad v=-a \exp\big[-{\rm i}(bx+cy+dt)\big],\quad Q=b^2-c^2+d,
\end{equation}
with $a,b,c,d\in \mathbb{R}$ and assume that the solution of the Lax pair
is in  the form of the following exponential function
\begin{equation}\label{fun2}
\begin{aligned}
&\psi=a_1\exp\big[{\rm i}(b_1x+c_1y+d_1t)\big],\\
&\phi=a_2\exp\big[{\rm i}(b_2x+c_2y+d_2t)\big],
\end{aligned}
\end{equation}
where $a_k\in \mathbb{R} $ and $b_{k},c_{k},d_{k}\in \mathbb{C}$ $(k=1,2)$.
Insertion of this expression into the Lax pair~(\ref{lax}), results
in the parameters of the above fundamental eigenfunction $\Psi=(\psi, \phi)^{T}$ (column vector solution to Lax pair) satisfying:
\begin{equation*}
\begin{aligned}
&a_2=\frac{(b_1+{\rm i}c_1)a_1}{a},\quad c_1^{\pm}=\frac{c+{\rm i}b}{2}\pm\frac{\Xi}{2},\\
&d_1^{\pm}=b^2-2bb_1+\frac{d}{2}\pm \frac{(-{\rm i}b+c+2{\rm i}b_1)\Xi}{2},\\
&b_2=b_1-b, \quad c_2=c_1-c, \quad d_2=d_1-d,\\
&\Xi=\sqrt{4a^2+c^2-(b-2b_1)^2+2{\rm i}c(b-2b_1)}.
\end{aligned}
\end{equation*}
For the sake of convenience, without loss of the generality of the possible dynamical behaviors for DS II equations, in what follows,
we always select $b=c=d=0$, in which case the seeding solution becomes
\begin{equation}\label{seed1}
u=-v=a,\quad Q=0,
\end{equation}
and the exponential eigenfunction (\ref{fun2}) reduces to
\begin{equation}\label{fun3}
\begin{aligned}
&\psi^{\pm}=\exp\big[{\rm i}b_1x\pm\sqrt{a^2-b_1^2}({\rm i}y-2b_1t)\big],\\
&\phi^{\pm}=\frac{b_1\pm\sqrt{a^2-b_1^2}{\rm i}}{a}\exp\big[{\rm i}b_1x\pm\sqrt{a^2-b_1^2}({\rm i}y-2b_1t)\big].
\end{aligned}
\end{equation}
 By performing a Taylor expansion for the above exponential
 eigenfunction  around the point $b_1=\lambda=\alpha+i\beta$ where $\alpha$ and $\beta$
are real constants and satisfy some constraints as seen in Remark \ref{rem0}, we have the power series:
\begin{equation}\label{taylor}
\begin{aligned}
&\psi(b_1=\lambda+\epsilon)=\psi^{[0]}+\psi^{[1]}\epsilon+\psi^{[2]}\epsilon^2+\cdots+\psi^{[N]}\epsilon^N+O(\epsilon^{N+1}),\\
&\phi(b_1=\lambda+\epsilon)=\phi^{[0]}+\phi^{[1]}\epsilon+\phi^{[2]}\epsilon^2+\cdots+\phi^{[N]}\epsilon^N+O(\epsilon^{N+1}),\\
\end{aligned}
\end{equation}
where $\psi^{[k]}=\frac{1}{k!}\frac{\partial^{k} \psi}{\partial
  b_1^k}|_{b_1=\lambda}, \phi^{[k]}=\frac{1}{k!}\frac{\partial^{k}
  \phi}{\partial b_1^k}|_{b_1=\lambda}, k=0,1,2,\cdots, N$ and $\epsilon>0$ is an
infinitesimal constant.
\begin{remark}\label{rem0}
To obtain higher-order lumps of the DS II equation, the parameters $\alpha$ and $\beta$ satisfy that $\alpha$ is arbitrary when $\beta\neq0$, or
$\alpha>a$ when $\beta=0$. In our paper, we focus on the case $\alpha=0,\beta\neq0$ below.
\end{remark} 
\begin{remark}\label{rem1}
  In what follows,  since the derivation of the eigenfunction components $\psi$ and $\phi$ with respect to
  the parameter variable $b_1$ results in the singularity of the denominator (the denominator shall contain $\sqrt{a^2-b_1^2}$), we avoid the degenerate case scenario by assuming
  hereafter that $\alpha=0,\beta\neq0$ .
\end{remark}

\begin{remark}\label{rem2}
Assume that $(\psi,\phi)^T$ solves the Lax pair (\ref{lax}) with the
seeding solution $u=-v=a, Q=0$. By
performing a Taylor expansion as in Eq.(\ref{taylor}),
the arbitrary order Taylor coefficients $(\psi^{[k]},\phi^{[k]})^T$ are solutions to Lax pair with $u=-v=a, Q=0$.
Based on the special seeding solution $u=-v=a$  (irrespectively of the expansion point $b_1\neq a$ as interpreted in Remark \ref{rem0})
and the analyticity of the eigenfunction $\Psi$ (see
Eq.~(\ref{fun3})), one can
conclude  that all derivatives of $\Psi$ with respect to variable $b_1$
satisfy the linear Lax pair Eq.~(\ref{lax}) with $u=-v=a$.
Its Taylor coefficients  $\Psi^{[k]}=(\psi^{[k]},\phi^{[k]})^T$ as above Eq.~(\ref{taylor}) are also
 solutions to the Lax pair  Eq.~(\ref{lax}) with the same seeding solution $u=-v=a$.
\end{remark}

\begin{remark}\label{rem3}
 We just consider the case of
$(\psi^+,\phi^{+})^T$, and for simplicity, we still use
$(\psi,\phi)^T$ instead of $(\psi^+,\phi^{+})^T$ below. For the case
with
$(\psi^-,\phi^{-})^T$ superscripts, the same dynamics of the solutions are obtained.
\end{remark}

Remark \ref{rem2} implies that there exists a hierarchy of eigenfunctions
composed of Taylor coefficients for the same seeding solution.
Based on the conclusion, the $n$th-order rational solution of the DS II equation
generated by the $n$-fold DT (Eq.~(49) in Ref.\cite{xinkou}) is generalized in the following Theorem.
\begin{thm}\label{thm2}
Given the seeding solution $u=-v=a$ and choosing $n$ distinct Taylor coefficients $\Psi^{[k_j]}=(\psi^{[k_j]},\phi^{[k_j]})^T$ ($k_j=1,2,\cdots,n$)
as eigenfunctions, then the new $n$th-order rational solution of DS II equation (\ref{DSII})
is given by
\begin{equation}\label{un}
u^{[n]}=a+2{\rm i}\frac{\delta_2}{\delta_1}
\end{equation}
where
\begin{equation*}
\begin{aligned}
\delta_1&=\begin{vmatrix}
\partial^{n-1}_x\psi^{[k_1]}_1 &\cdots&\partial^{n-1}_x\psi^{[k_n]}_n &\partial^{n-1}_x\phi^{[k_1]*}_1 &\cdots &\partial^{n-1}_x\phi^{[k_n]*}_n\\
\partial^{n-2}_x\psi^{[k_1]}_1 &\cdots&\partial^{n-2}_x\psi^{[k_n]}_n &\partial^{n-2}_x\phi^{[k_1]*}_1 &\cdots &\partial^{n-2}_x\phi^{[k_n]*}_n\\
\vdots&\vdots&\vdots&\vdots&\vdots&\vdots\\
\psi^{[k_1]}_1 &\cdots &\psi^{[k_n]}_n &\phi^{[k_1]*}_1 &\cdots &\phi^{[k_n]*}_n\\
\partial^{n-1}_x\phi^{[k_1]}_1 &\cdots&\partial^{n-1}_x\phi^{[k_n]}_n &-\partial^{n-1}_x\psi^{[k_1]*}_1 &\cdots &-\partial^{n-1}_x\psi^{[k_n]*}_n\\
\partial^{n-2}_x\phi^{[k_1]}_1 &\cdots&\partial^{n-2}_x\phi^{[k_n]}_n &-\partial^{n-2}_x\psi^{[k_1]*}_1 &\cdots &-\partial^{n-2}_x\psi^{[k_n]*}_n\\
\vdots&\vdots&\vdots&\vdots&\vdots&\vdots\\
\phi^{[k_1]}_1 &\cdots &\phi^{[k_n]}_n &-\psi^{[k_1]*}_1 &\cdots &-\psi^{[k_n]*}_n
\end{vmatrix}
\end{aligned} \nonumber
\end{equation*}
and $\delta_2$ is the $n+1$ row of $\delta_1$ replaced by a row vector $\eta$=$(\partial^{n}_x\psi^{[k_1]}_1 ,\cdots,\partial^{n}_x\psi^{[k_n]}_n ,\partial^{n}_x\phi^{[k_1]*}_1 ,\cdots \partial^{n}_x\phi^{[k_n]*}_n)$.
\end{thm}

\begin{remark}
Comparing with the $n$-fold DT (see Eq.(49) in Ref.\cite{xinkou}), we
use
here the Taylor coefficients as
new eigenfunctions in order to construct a variety of solutions of DS II.
\end{remark}

\section{The higher-order lump solutions}
For the DS II equations, under certain parameter restrictions,
multi-lump solutions have been obtained in \cite{JMP20} on nonzero
background. Later, M. Ma\~nas {\it et al.}
employed a Wronskian scheme and  M. Ablowitz {\it et al.} employed the
inverse scattering transformation and Laurent coefficients to study
the lump solutions on
top of a zero background\cite{PLA1997,SIAM2003}. They have found that
some lumps described a non-trivial interaction, in other words, after a
front collision, these lumps underwent scattering with
a $\frac{\pi}{2}$ scattering
angle.  In the current work, we shall investigate higher-order lumps
of the DS II equation
on top of a nonzero constant background and  show that they
possess novel
scattering where the scattering angle can be in the interval of
$(\frac{\pi}{2},\pi)$.

\begin{figure}[!htbp]
\centering
\raisebox{19 ex}{$|u^{[1]}|$}\subfigure[]{\includegraphics[height=4.5cm,width=4.8cm]{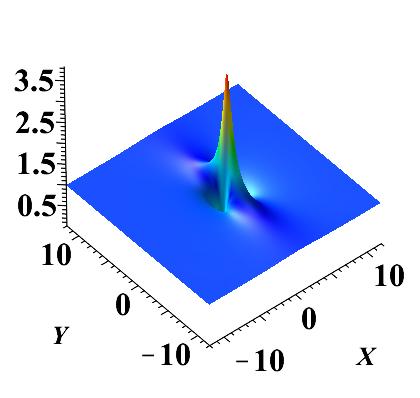}}\quad
\raisebox{16 ex}{}\subfigure[]{\includegraphics[height=4.5cm,width=4.8cm]{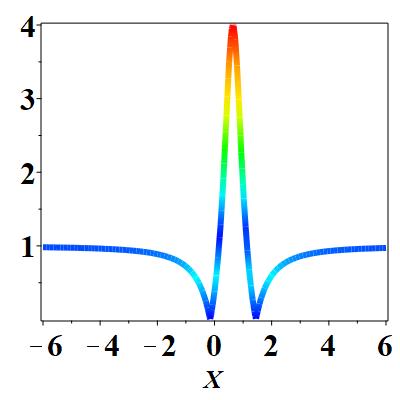}}\quad
\caption{(Color online) (a) The first-order lump of the DS II equation with parameters $a=1,
\beta=\frac{1}{2}, M=\frac{\sqrt{5}}{2}$, (b) the $Y$-crossection showing one maximum and two minima.
  }\label{1lump}
\end{figure}

\noindent{\bf A: the first-order fundamental lump}$\\$
\indent In this part,
without loss of diversity of dynamical behaviors of lump solutions, for simplicity,
we choose the pure imaginary expanding point $\lambda={\rm i}\beta$ in Eq.~(\ref{taylor}), i.e., $\alpha=0$,
and consider the
following new moving coordinate frame
\begin{equation}\label{moving1}
X=x, \quad Y=y-(4\beta+\frac{2a^2}{\beta})t.
\end{equation}
The two eigenfunction components are given by the first-order Taylor coefficients
\begin{equation}
\begin{aligned}
\psi_1=\psi^{[1]}=&\frac{({\rm i}MX+\beta Y)e^{\frac{{\rm i}({\rm i}\beta^2 X+2M^3t+\beta MY)}{\beta}}}{M},\\
\phi_1=\phi^{[1]}=&\frac{(M+\beta)(1-MX+{\rm i}\beta Y)e^{\frac{{\rm i}({\rm i}\beta^2 X+2M^3t+\beta MY)}{\beta}}}{M a},
\end{aligned}
\end{equation}
with $M=\sqrt{a^2+\beta^2}$.
Then, insertion of $(\psi_1,\phi_1)^T=(\psi^{[1]},\phi^{[1]})$ into
the one-fold DT, Eq. (\ref{un}) with parameters $n=1, k_1=1$ yields a first-order fundamental lump solution
\begin{equation}
u^{[1]}=a\Big[1+\frac{-1+2{\rm i}\beta Y}{\big(MX-\frac{M+\beta}{2M}\big)^2+\frac{a^2}{4M^2}+\beta^2Y^2}\Big].
\end{equation}
It was first obtained by Satsuma and Ablowitz by taking a ``long wave" limit of the corresponding one-soliton solution
constructed by the direct method~\cite{JMP20}.
 The solution is stationary in the moving coordinate $(X, Y)$- frame. It has a single maximum peak $\frac{(3M^2+\beta^2)}{a^2}|a|$ at ($\frac{M+\beta}{2M^2}, 0$) and two local minima $0$
at ($\frac{M+\beta}{2M^2}\pm\frac{\sqrt{3M^2+\beta^2}}{2M^2}, 0$).
Recalling the transformation connecting the moving
frame to the rest one ($(x,y)$-plane), this first-order lump travels with a uniform velocity $(0, 4\beta+\frac{2a^2}{\beta})$.
 Its dynamics is illustrated in the $(X,Y)$-plane in Fig.~\ref{1lump}.

\begin{remark}
  In the above section, when $\lambda$ is real and $|\lambda|\leq|a|$ ,
  we observe that the line rogue wave solutions of DS II are
  obtained. On the other hand, we note that
when $\lambda$ is a complex constant (or pure imaginary constant) and $|\lambda_{\Re}|>|a|$ (or $\lambda={\rm i} \beta$) where
$\lambda_{\Re}$ denotes the real part of $\lambda$, the lump solutions
of DS II are derived, i.e., the character of the solutions changes
depending
on the specific selection of  $\lambda$ within the complex plane.

\end{remark}
\begin{remark}
In what follows, we shall set $a>0$ and $\beta>0$ without loss of generality.
\end{remark}
\noindent{\bf B: the second-order non-fundamental lump}$\\$
\indent This part is devoted to using one- and two-fold DT to
construct two
second-order non-fundamental lump solutions and study
their dynamical properties.
For the sake of convenience, the discussion is considered in the moving coordinate frame (\ref{moving1}) below.\\
{\it Case 1 the second-order lump using one-fold DT}

   In this case, choosing the following set of parameters in Eq. (\ref{un})
   \begin{equation}
   \lambda_1={\rm i}\beta(\beta>0), \quad n=1, \quad k_1=2,
   \end{equation}
   then the one-fold DT yields a second-order non-fundamental lump solution.
   Because of the cumbersome expression of this solution, we just
   provide
   here the corresponding eigenfunctions
  \begin{equation}\label{fun}
  \begin{aligned}
  \psi^{[2]}=&\frac{{\rm i}(2M^2\beta^2XY-2M^4t+6M^2\beta^2t-M^2\beta Y+\beta^3Y+{\rm i}M^3\beta^2X^2-{\rm i}M\beta^3Y^2)e^{-\beta^2X+{\rm i}(2M^3t+M\beta Y)}}{2M^3\beta},\\
  \phi^{[2]}=&\frac{(M+\beta){\rm i}}{a}\psi^{[2]}+\frac{\big[{\rm i}M^2X+(\beta Y-\frac{{\rm i}}{2})M+\frac{{\rm i}\beta}{2}\big](M+\beta)e^{-\beta^2X+{\rm i}(2M^3t+M\beta Y)}}{M^3\beta}.
  \end{aligned}
  \end{equation}
  Distinctly from the first-order case, the two
    eigenfunction components are dependent on $t$,
  so the solution is non-stationary in the moving coordinate frame. When $|t|\rightarrow\infty$, it contains two separated individual
  lump peaks, while in the intermediate times, the two lump peaks fuse together. In order to analyze their interaction process,
  their heights and the traveling paths of the two local maxima for
  the two lump peaks
  need to be determined.
  Since the exact analytical formulas are very complicated to obtain, we make the following reductions.
  Taking into consideration the expression
  \begin{equation}\label{u2}
  u^{[2]}=a-2{\rm i}\frac{\psi^{[2]}\phi^{[2]*}_X-\phi^{[2]}\psi^{[2]*}_X}{|\psi^{[2]}|^2+|\phi^{[2]}|^2},
  \end{equation}
  the maximum of $|u^{[2]}|$ shall occur near the
  minimum of the denominator which is approximately at the zeros of the leading terms of this polynomial part.

\begin{figure}[!htbp]
\centering
\raisebox{16 ex}{}{\includegraphics[height=4.5cm,width=4.8cm]{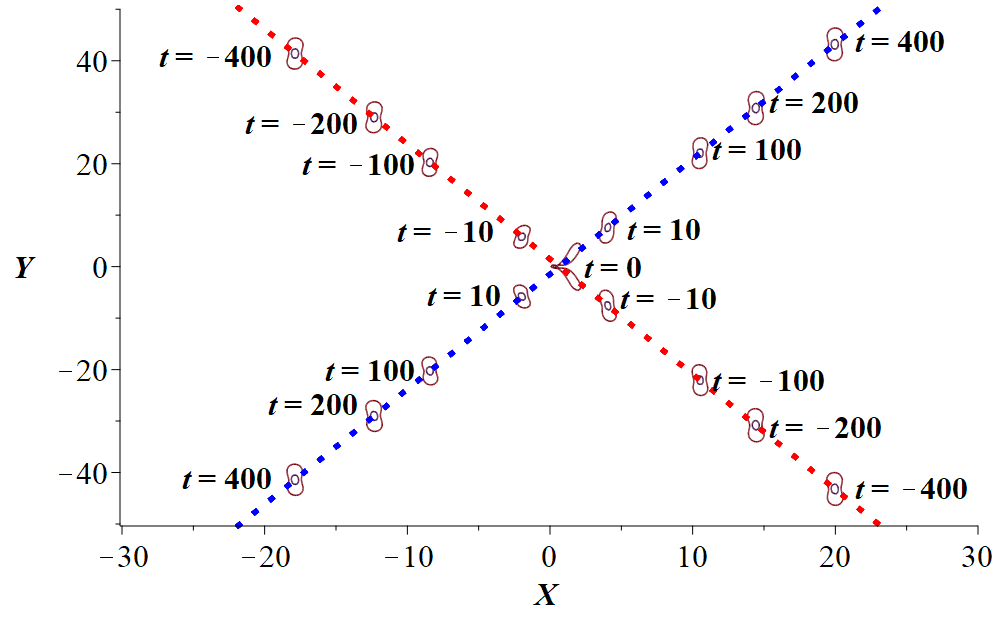}}\quad
\caption{(Color online) The time evolution of the second-order lump of
  Eq.~(\ref{u2}). The contour plots of these lumps at distinct times
  is plotted
  using the exact analytical solution (\ref{u2}) by leveraging the
  eigenfunction of Eq.~(\ref{fun}), and the two dot straight lines
  denote the two asymptotic
  lines given by Eqs.~(\ref{twolumpasy1}) and (\ref{twolumpasy2}).}\label{numerlump}
\end{figure}

  Solving $|\psi^{[2]}|=0$,
  when $\beta>0, M^2-3\beta^2=a^2-2\beta^2>0$, for large time $|t|$,
  we identify the two lump peaks whose maximum asymptotic coordinates are given by
  \begin{equation}\label{twolumpasy1}
  X=\pm\frac{\sqrt{-M\beta (M^2-3\beta^2)t}}{M\beta}+\frac{a^2}{4\beta M^2}+\frac{M+\beta}{2M^2},\quad Y=-\frac{M}{\beta}X+\frac{M+\beta}{2M\beta},\quad t\rightarrow-\infty,
  \end{equation}
  and
  \begin{equation}\label{twolumpasy2}
  X=\pm\frac{\sqrt{M\beta(M^2-3\beta^2)t}}{M\beta}+\frac{a^2}{4\beta M^2}+\frac{M+\beta}{2M^2},\quad Y=\frac{M}{\beta}X-\frac{M+\beta}{2M\beta},\quad
  t\rightarrow+\infty,
  \end{equation}
  with  $M=\sqrt{a^2+\beta^2}$.
  It is confirmed by direct comparison with the numerical profiles
    of the exact solutions that these approximate asymptotic
    expressions
    adequately reflect the
 lump center positions for all times as shown in Fig.~\ref{numerlump}.

\begin{figure}[!htbp]
\centering
\raisebox{16 ex}{}\subfigure[]{\includegraphics[height=4.5cm,width=4.8cm]{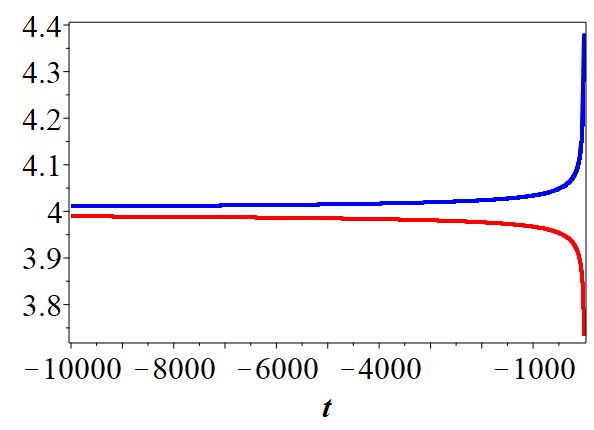}}\quad
\raisebox{16 ex}{}\subfigure[]{\includegraphics[height=4.5cm,width=4.8cm]{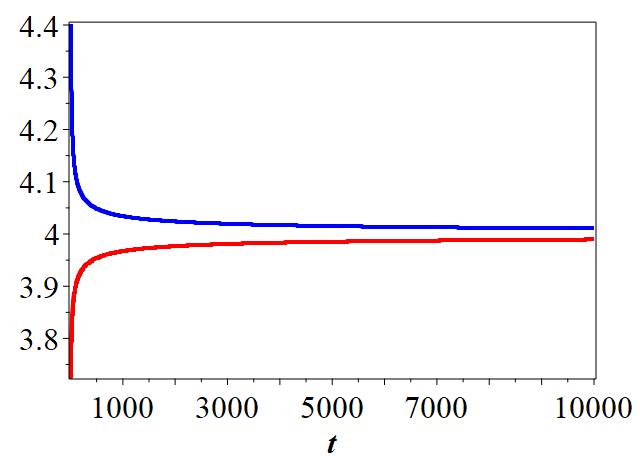}}\quad
\caption{(Color online) (a) the two maximum amplitude values of the
  two lumps from $t=-10000$ to $t=-10$, (b) the two maximum amplitude
  values of
  the two lumps from $t=10$ to $t=10000$.}\label{amplitude}
\end{figure}

  Before studying the  maximum amplitudes of the two lumps, we take
  $a=1, \beta=\frac{1}{2}, M=\frac{\sqrt{5}}{2}$.  Substituting the asymptotic
  coordinates (\ref{twolumpasy1}) and (\ref{twolumpasy2}) into the solution $u^{[2]}$ (\ref{u2}), the amplitude of one lump is
  larger than $4$ and approaches $4$, whereas for the other one, the
  amplitude is less
  than $4$ and approaches $4$ as
  $t\rightarrow-\infty$. The same conclusion holds
  as $t\rightarrow+\infty$, as shown in Fig.~\ref{amplitude}.
 Notice there is a slight deviation of our above asymptotic
  expressions from the limiting value, given their approximate nature.


\begin{figure}[!htbp]
\centering
\raisebox{16 ex}{}\subfigure[]{\includegraphics[height=4.5cm,width=5.5cm]{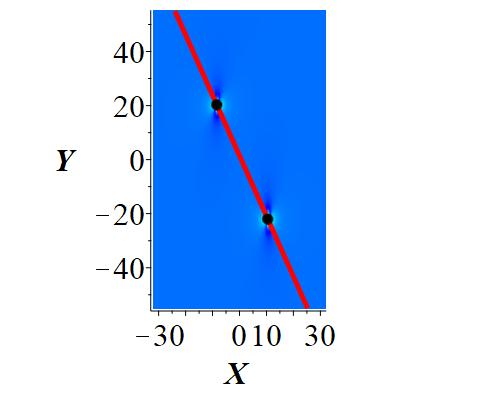}}
\raisebox{16 ex}{}\subfigure[]{\includegraphics[height=4.5cm,width=5.5cm]{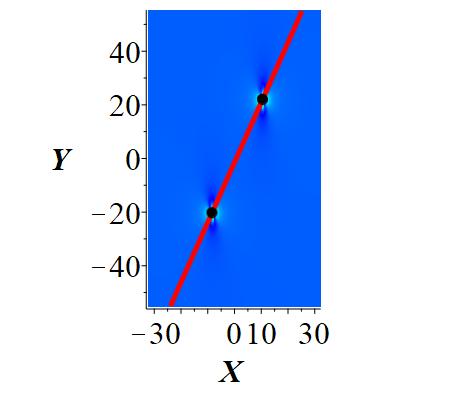}}
\raisebox{16 ex}{}\subfigure[]{\includegraphics[height=4.5cm,width=5.5cm]{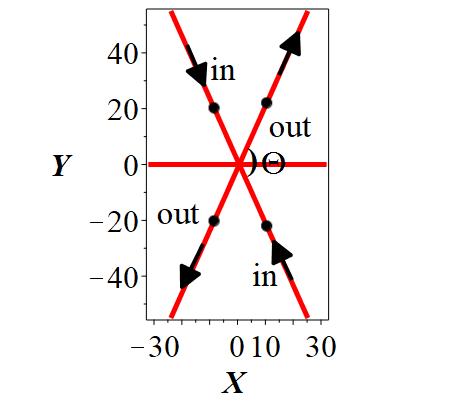}}
\caption{(Color online) Location of the two pulses: (a) incoming ($t=-100$)  and (b) outgoing ($t=100$) with parameters $a=1,\beta=\frac{1}{2}, M=\frac{\sqrt{5}}{2}$.
(c) The nontrivial collision process of before ($t \rightarrow
-\infty$) and after
($t \rightarrow +\infty$) scattering and the angle is indicated, and
when $|t|=100$, these two lumps nearly totally locate at the two
straight lines. In panel (c), a schematic of the incoming and outgoing wave
angles
is provided.
}\label{display}
\end{figure}

From Eqs.~(\ref{twolumpasy1}) and (\ref{twolumpasy2}), the asymptotic trajectories define
two straight lines with different slopes for $t\rightarrow\pm\infty$. More concretely, the  asymptotic line for $t \rightarrow +\infty$
is obtained from  the  line for $t \rightarrow -\infty$  by reflection
with respect to the $X$-axis, and the angle between the
two asymptotic lines is denoted by $\Theta$.
Since the two lump peaks first experience a head-on collision henceforth undergoing a scattering
process, according to the coordinates (\ref{twolumpasy1}) and (\ref{twolumpasy2}),
the scattering angle $\Theta$ is given by
  \begin{equation}\label{scatteringangle1}
  \cos\Theta=-\frac{a^2}{a^2+2\beta^2}.
  \end{equation}
  Here $\cos \Theta$ reaches to the minimum value $-1$  as $a\rightarrow\infty$ and attains a maximum value $0$ as $a$ goes to $0$; in other words,
  the scattering angle $\Theta\in(\frac{\pi}{2},\pi)$. Fig.~\ref{display} shows the traveling paths of the two lump peaks before and after collision. It is
  seen that the two lumps located at the second and fourth quadrants first accelerate and approach each other
  along a straight line. After a front collision and undergoing
  a large scattering angle, they decelerate and move away each other
  along the other straight line (among the ones given above) and,
  finally, they move along the first and third quadrants.
  Also, the approximate estimations of the center positions are
  nearly coincident with the exact ones illustrated by the density plots.

\begin{figure}[!htbp]
\centering
\raisebox{16 ex}{\includegraphics[height=4.5cm,width=4.8cm]{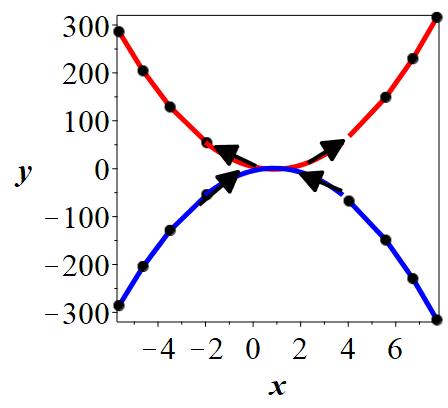}}\quad
\caption{(Color online) The second-order lump peak trajectories with parameters $a=1, \beta=\frac{1}{2}$ and
$M=\frac{\sqrt{5}}{2}$ in the $xy$-plane. The blue parabolic curve
is given by the first formula of Eqs.~(\ref{resttraj}) and (\ref{time}) , the red parabolic curve
is given by the second formula of Eqs.~(\ref{resttraj}) and (\ref{time}), and the black
points describe the peak coordinates.}\label{restframe}
\end{figure}

Reverting back to the rest coordinate $(x,y)$- frame, from Eq.~(\ref{moving1}), the peak locations
are given by $(x,y)=(X, \frac{2a^2t+4\beta^2t+\beta Y}{\beta})$ when $|t|\gg0$. The corresponding coordinate $y$ satisfies
 the following equations,
\begin{equation}\label{resttraj}
\begin{aligned}
\beta y=&(2a^2+4\beta^2)t-Mx+\frac{M+\beta}{2M},\qquad t\rightarrow-\infty,\\
\beta y=&(2a^2+4\beta^2)t+Mx-\frac{M+\beta}{2M}, \qquad t\rightarrow+\infty,\\
\end{aligned}
\end{equation}
where $t$ can given by solving Eqs.~(\ref{twolumpasy1}) and (\ref{twolumpasy2}), i.e.,
\begin{equation}\label{time}
|t|=\frac{\sqrt{M\beta}}{M^2-3\beta^2}\Big[x-\frac{a^2+2\beta(M+\beta)}{4\beta M^2}\Big]^2
\end{equation}
with $M>0, \beta>0$ and $M^2-3\beta^2>0$ when $a^2>2\beta^2$. Combining Eqs.(\ref{resttraj}) and (\ref{time}), it is found
that the two lumps locate at two parabolas in the $(x,y)$-plane, which is illustrated in Fig.~\ref{restframe}.

\begin{figure}[!htbp]
\centering
\raisebox{16 ex}{}\subfigure[]{\includegraphics[height=4.5cm,width=5.3cm]{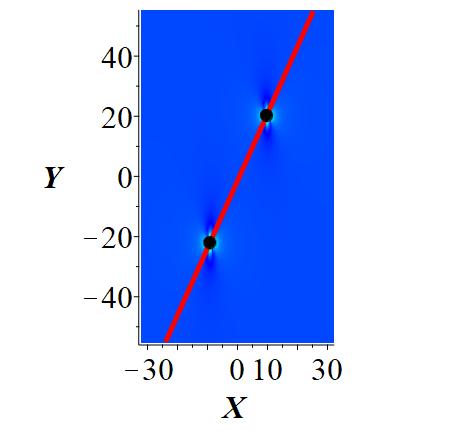}}
\raisebox{16 ex}{}\subfigure[]{\includegraphics[height=4.5cm,width=5.5cm]{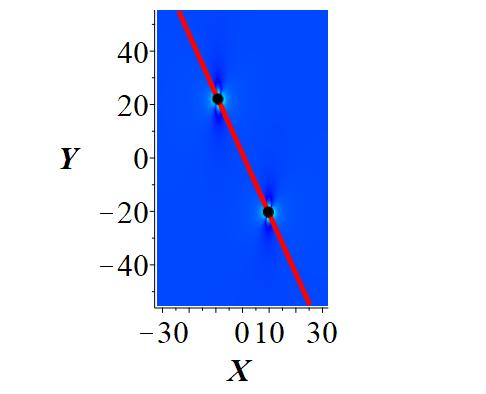}}
\raisebox{16 ex}{}\subfigure[]{\includegraphics[height=4.5cm,width=5.3cm]{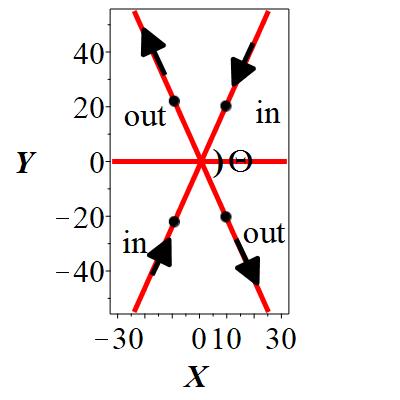}}
\caption{(Color online) Location of the two pulses: (a) incoming ($t=-100$)  and (b) outgoing ($t=100$) with parameters $a=1,\beta=\frac{1}{2},M=\frac{\sqrt{5}}{2}$.
(c) The nontrivial collision process of before ($t=-\infty$) and after
($t=+\infty$) scattering and the angle is indicated. When $|t|=100$,
these two lumps are practically located at two straight lines. The scattering angle is also indicted by $\Theta$.}\label{display3}
\end{figure}

\noindent {\it Case 2 the second-order lump using two-fold DT}$\\$
\indent To compare with the second-order lump in Case 1, we choose the following set of parameters in Eq. (\ref{un})
\begin{equation}
\lambda_2=\lambda_1={\rm i}\beta, \quad n=2, \quad k_1=1, \quad k_2=2.
\end{equation}
A new second-order lump is obtained by using the two-fold DT. For large time $t$,
this solution features  a generally opposite time evolution process in
comparison
to the  Case 1 above, {that is, when $t\rightarrow-\infty$ the two lump
peaks locate at the first and third quadrants whereas they move to the second and
fourth quadrants as  $t\rightarrow+\infty$.
 To demonstrate this phenomenon, the asymptotical trajectories of these two lump
peaks are determined.}
Similarly to our discussion above,
the approximate coordinates of the maxima of two lumps are given by
\begin{equation}\label{twolumpasy3}
X=\pm\frac{\sqrt{-M\beta(M-3\beta^2)t}}{M\beta}-\frac{a^2}{4M^2\beta}+\frac{M+\beta}{2M^2},\quad
Y=\frac{M}{\beta}X-\frac{M+\beta}{2M\beta},\quad
t\rightarrow-\infty,
\end{equation}
and
\begin{equation}\label{twolumpasy4}
X=\pm\frac{\sqrt{M\beta(M^2-3\beta^2)t}}{M\beta}-\frac{a^2}{4M^2\beta}+\frac{M+\beta}{2M^2},\quad
Y=-\frac{M}{\beta}X+\frac{M+\beta}{2M\beta},
\quad t\rightarrow+\infty.
\end{equation}
where $M=\sqrt{a^2+\beta^2}$ and $\sqrt{M\beta(M^2-3\beta^2)|t|}$ is well-defined as $a^2>2\beta^2$. In this case, the scattering angle is given by
\begin{equation}\label{scatteringangle2}
\cos\theta=-\frac{a^2}{a^2+2\beta^2}
\end{equation}
This asymptotic dynamics is illustrated in Fig.~\ref{display3}.  It is clearly seen that these approximate estimations are
in good agreement with the exact solution illustrated by the density plot. Furthermore, comparing Fig.~\ref{display3}(c)
with Fig.~\ref{display}(c), it is also found that the second-order
lump obtained by using the one-fold DT evolves effectively in a time-reversed
way in comparison with the one obtained by using two-fold DT (see also Eqs.(\ref{twolumpasy1}) and (\ref{twolumpasy4})
and Eqs.(\ref{twolumpasy2}) and (\ref{twolumpasy3}), respectively).
\\
\begin{remark}
Though the two scattering angles in Case 1 and Case 2 are the same (see Eqs.(\ref{scatteringangle1}) and (\ref{scatteringangle2})),
the directions of incoming and outgoing waves are opposite.
Also,
the scattering angle is not necessarily normal, which is
a central difference of the results herein from the one of two lumps for DS II equation on zero background~\cite{PLA1997,SIAM2003}.
\end{remark}

\noindent{\bf C: the non-fundamental third-order lump}\\
{The third-order lump and its asymptotics can be studied in the same manner as in the
  second-order lump case,  and for this reason we omit here some  of the
  technical details. To illustrate its dynamical
evolution process and asymptotic heights, the locations of the three lump peaks shall
be given.}\\
\noindent {\it Case 3 the third-order lump using the one-fold DT}$\\$
 With the choice of the following set of parameters in Eq. (\ref{un}):
\begin{equation}\label{para1}
\lambda_1={\rm i}\beta, \quad n=1, \quad k_1=3,
\end{equation}
a third-order lump $u^{[3]}$ is obtained by using one-fold DT.
The following eigenfunctions are used,
\begin{equation}
\psi^{[3]}=-\frac{1}{6M^5\beta}(\psi_{\rm Re}^{[3]}+{\rm i}\psi_{\rm Im}^{[3]})e^{\xi},\quad
\phi^{[3]}=\frac{-{\rm i}(M+\beta)}{M^5\beta a}(\phi_{\rm Re}^{[3]}+{\rm i}\phi_{\rm Im}^{[3]})e^{\xi},
\end{equation}
with
\begin{equation*}
\begin{aligned}
\xi&=-\beta^2X+{\rm i}(M\beta Y+2M^3t),\quad M=\sqrt{a^2+\beta^2},\\
\psi_{\rm Re}^{[3]}&=3\beta^2M^4X^2Y-\beta^4M^2Y^3-6M^6tX+18M^4\beta^2tX-3\beta M^4XY+3\beta^3M^2XY\\
&\ \ \ -12\beta M^4t+12\beta^3M^2t-3\beta^2M^2Y+3\beta^4Y,\\
\psi_{\rm Im}^{[3]}&=M\beta(M^4X^3-3\beta^2M^2XY^2+6M^4tY-18\beta^2M^2tY+3M^2\beta Y^2-3\beta^3Y^2),\\
\phi_{\rm Re}^{[3]}&=\psi_{\rm Re}^{[3]}-6\beta^2M^3XY+6M^5t-18M^3\beta^2t+3\beta M^3Y+3\beta^2M^2Y-6\beta^3MY,\\
\phi_{\rm Im}^{[3]}&=\psi_{\rm Im}^{[3]}-3\beta M^4X^2+3M^2\beta^3Y^2+3M^3\beta X-3\beta^2M^2X+3M\beta^2-3\beta^3.
\end{aligned}
\end{equation*}
{For large $t$, this solution is split into three lumps whose
asymptotic coordinates of
the maxima are given by
\begin{equation}\label{3lumpasytraj1}
\left\{\begin{array}{c}
\small{\begin{aligned}
&X_{1,2}=\pm\frac{\sqrt{3M\beta(M^2-3\beta^2)t}}{M\beta}+\Delta+\frac{M-\beta}{3M\beta},
\quad Y=-\frac{M}{\beta}X_{1,2}+\frac{M+\beta}{M\beta}+\frac{a^2}{4M\beta},\\
&X_3=-\frac{M+\beta}{3M},\quad Y_3=-\frac{M}{\beta}X_3-\frac{M+\beta}{2M\beta}-\frac{a^2}{3M\beta},
\end{aligned}}
\end{array}
\right. \qquad t\rightarrow-\infty,
\end{equation}
and
\begin{equation}\label{3lumpasytraj2}
\left\{\begin{array}{c}
\small{\begin{aligned}
&X_{1,2}=\pm\frac{\sqrt{-3M\beta(M^2-3\beta^2)t}}{M\beta}+\Delta+\frac{M-\beta}{3M\beta},
\quad Y=\frac{M}{\beta}X_{1,2}-\frac{M+\beta}{M\beta}-\frac{a^2}{4M\beta},\\
&X_3=-\frac{M+\beta}{3M},\quad Y_3=\frac{M}{\beta}X_3+\frac{M+\beta}{2M\beta}+\frac{a^2}{3M\beta},
\end{aligned}}
\end{array}
\right. \qquad t\rightarrow+\infty,
\end{equation}
where $\Delta=\frac{3(M+\beta)(M\beta+M-\beta)}{4M^2\beta}$ and when
$a^2>2\beta^2$ the quantity $\sqrt{3M\beta(M^2-3\beta^2)|t|}$ is well defined.
The dynamics of this third-order lump is illustrated in Fig.~\ref{3lump}. Fig.~\ref{3lumpasy} shows that the exact (density figures) and approximate
peak locations (red, blue and black points) are generally in good agreement for large $|t|$. When $t\ll0$, the three lump peaks
are separated and two of them are located at the second and fourth
quadrants.
Subsequently they approach and eventually overlap with the middle one.
As time progresses, the three lump peaks again split into three
distinguishable peaks, with the middle one remaining fixed while the
other two peaks separate from each other and move to the first and
third quadrants.
Note that the two peaks (located at the
first and third quadrants (or the second and fourth quadrants) move along two distinct straight lines but their slopes are same with the
corresponding second-order lump obtained by using one-fold DT (see
also
Eqs.~(\ref{twolumpasy1})--(\ref{twolumpasy2}) and (\ref{3lumpasytraj1})--(\ref{3lumpasytraj2})).
\begin{figure}[!htbp]
\centering
\raisebox{16 ex}{}\subfigure[]{\includegraphics[height=3.5cm,width=3.5cm]{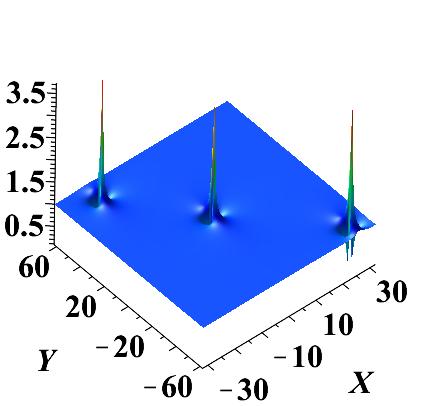}}
\raisebox{16 ex}{}\subfigure[]{\includegraphics[height=3.5cm,width=3.5cm]{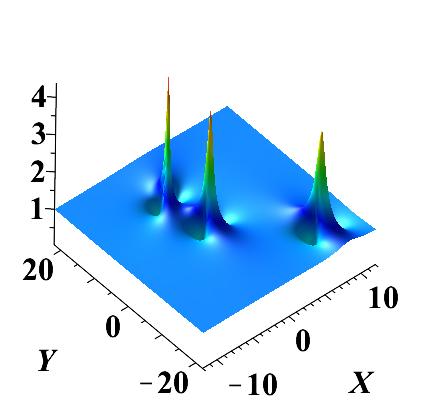}}
\raisebox{16 ex}{}\subfigure[]{\includegraphics[height=3.5cm,width=3.5cm]{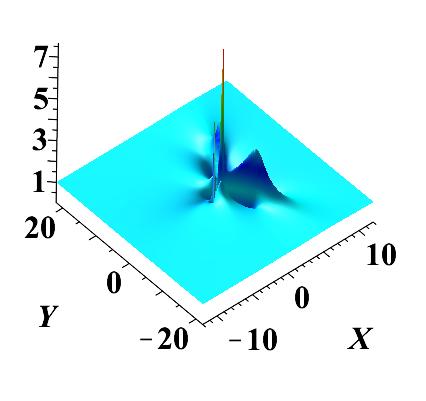}}
\raisebox{16 ex}{}\subfigure[]{\includegraphics[height=3.5cm,width=3.5cm]{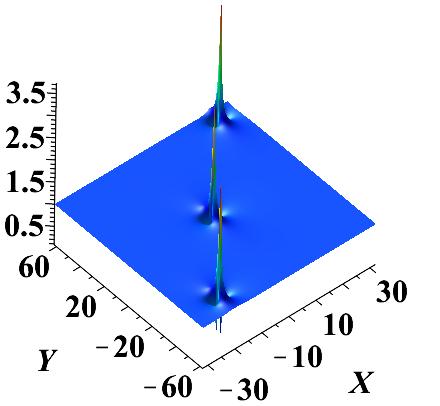}}
\caption{(Color online) The time evolution process of the third-order
  lump obtained by using the one-fold
DT with parameters $a=1, \beta=\frac{1}{2}$ and $M=\frac{\sqrt{5}}{2}$ at distinct time. (a) $t=-200$; (b) $t=-10$; (c) $t=0$; (d) $t=200$.
}\label{3lump}
\end{figure}
\begin{figure}[!htbp]
\centering
\raisebox{16 ex}{}\subfigure[]{\includegraphics[height=4.5cm,width=4.8cm]{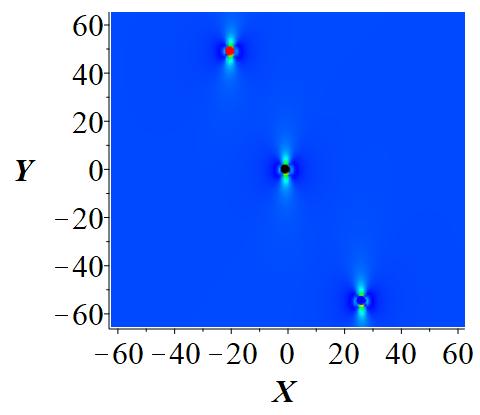}}\qquad
\raisebox{16 ex}{}\subfigure[]{\includegraphics[height=4.5cm,width=4.8cm]{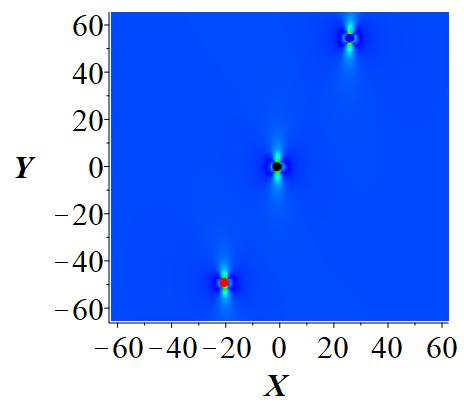}}\quad
\caption{(Color online) Location of the three lump peaks in Case 3: (a) $t=-200$, (b) $t=200$.
 The black point represents the location of the fixed lump; the red
 point denotes the approximate coordinates of one lump, and the blue point denotes the approximate
 coordinates of the other  lump.
}\label{3lumpasy}
\end{figure}
Furthermore, the approximate heights of the three lump peaks are also calculated by
substituting the asymptotic coordinates into the expression of the third-order lump solution,
as illustrated in Fig.~\ref{3lumpheight}. It is seen that (i) each peak height approaches
the asymptotic value $4$ where the minor difference comes from the
approximate coordinate estimate (similarly to what was discussed before)
of lump peaks; (ii) the peak height (red point) grows
as time evolves whereas the other one (blue point) decreases as $t\ll0$, but the peak height (red point) decreases
as time evolves whereas the other one (blue point) grows for $t\gg0$. Moreover, the middle one (black point)
generally remains unchanged during the evolution process.

\begin{figure}[!htbp]
\centering
\raisebox{16 ex}{}\subfigure[]{\includegraphics[height=4.5cm,width=4.8cm]{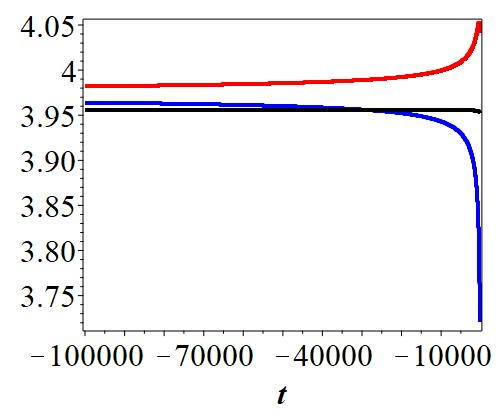}}\qquad
\raisebox{16 ex}{}\subfigure[]{\includegraphics[height=4.5cm,width=4.8cm]{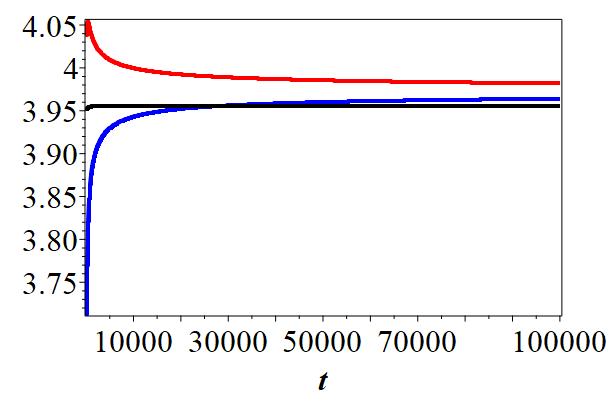}}\quad
\caption{(Color online) The evolution of the heights of the three lump
  peaks
  over time: (a) from $t=-100000$ to $t=-200$,
(b) from $t=200$ to $t=100000$.
 The black line represents the height of the fixed lump; the red
 line denotes the height of second lump, and the blue line denotes the
 height of
 the third  lump.
}\label{3lumpheight}
\end{figure}

}

\begin{remark}
The trajectories of the three lump peaks in the $xy$-plane can be obtained using (\ref{moving1}), (\ref{3lumpasytraj1})
and (\ref{3lumpasytraj2}) as in the two-lump case but are not shown here.
\end{remark}

\begin{figure}[!htbp]
\centering
\raisebox{16 ex}{}\subfigure[]{\includegraphics[height=4.5cm,width=4.8cm]{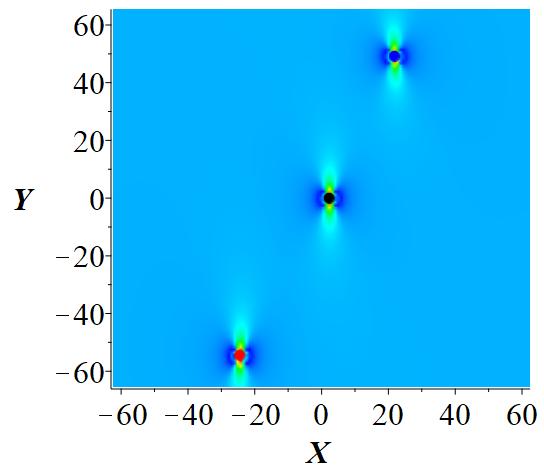}}\qquad
\raisebox{16 ex}{}\subfigure[]{\includegraphics[height=4.5cm,width=4.8cm]{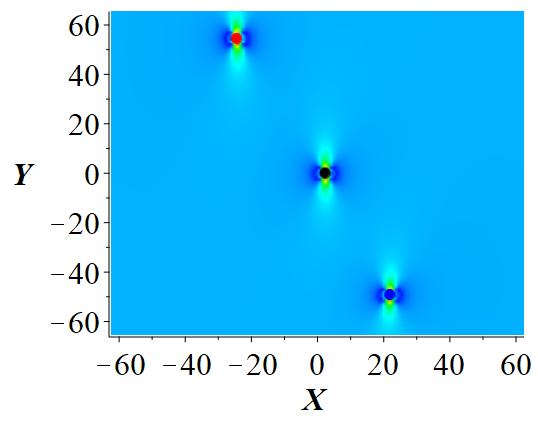}}\quad
\caption{(Color online) Location of the three lump peaks in Case 4: (a) $t=-200$, (b) $t=200$.
 The black point represents the location of the fixed lump; the red
 points denote the approximate coordinates of the second lump, and the
 blue points
 denote the approximate
 coordinates of the third  lump.
}\label{3lumpasy1}
\end{figure}

\noindent {\it Case 4 the third-order lump using three-fold DT}$\\$
\indent {Comparing with the third-order lump obtained by using the
  one-fold DT in Case 3, in
the present case, we shall use the three-fold DT to construct a similar third-order lump,
but which possesses a generally ``opposite'' time evolution
process. That is, when $t\ll0$ two lumps are located
at the first and third quadrants whereas they move to the second and fourth quadrants as
$t\gg0$, with the middle lump remaining still during the entire time evolution.}\\
\indent Choosing the following set of parameters in Eq. (\ref{un})
\begin{equation}\label{para2}
\lambda_1=\lambda_2=\lambda_3={\rm i}\beta,  \quad n=3, \quad k_1=1, \quad k_2=2,\quad k_3=3,
\end{equation}
a third-order lump $\widetilde{u^{[3]}_{lump}}$ of DS II is obtained by using three-fold DT. {Since the
  expression of this solution is lengthy and complex, once again
  we leverage the analytical means of approximating the trajectories and
  heights of the three lump peaks similarly to previous cases.
  Indeed, we omit lengthy details but only focus on some relevant
  results for
  the time evolution of the lump peaks.}
For large $t$, the approximate coordinates of
these three lumps are given by
\begin{equation}\label{3lumpasytraj3}
\left\{\begin{array}{c}
\begin{aligned}
X_{1,2}&=\pm\frac{\sqrt{-3M\beta(S^2-3\beta^2)t}}{M\beta}-\Delta+\frac{M-\beta}{M^2\beta},\quad Y=\frac{M}{\beta}X_{1,2}+\frac{M+\beta}{M\beta}-\frac{3a^2}{2M\beta},\\
X_3&=\Delta,\quad y=\frac{M}{\beta}X_3-\frac{2(M+\beta)}{M^2\beta},
\end{aligned}
\end{array}
\right. \qquad t\rightarrow-\infty,
\end{equation}
and
\begin{equation}\label{3lumpasytraj4}
\left\{\begin{array}{c}
\begin{aligned}
X_{1,2}&=\pm\frac{\sqrt{3M\beta(S^2-3\beta^2)t}}{M\beta}-\Delta+\frac{M-\beta}{M^2\beta},\quad Y=-\frac{M}{\beta}X_{1,2}-\frac{M+\beta}{M\beta}+\frac{3a^2}{2M\beta},\\
X_3&=\Delta,\quad y=-\frac{M}{\beta}X_3+\frac{2(M+\beta)}{M^2\beta},
\end{aligned}
\end{array}
\right. \qquad t\rightarrow+\infty,
\end{equation}
where $\Delta=\frac{3(M+\beta)(M\beta+M-\beta)}{4M^2\beta}$
and when $a^2>2\beta^2$ the quantity
$\sqrt{3M\beta(M^2-3\beta^2)|t|}$ is well defined.
When $|t|=200$ the exact analytical solution and approximate coordinates (\ref{3lumpasytraj3})
and (\ref{3lumpasytraj4}) are plotted in Fig.~{\ref{3lumpasy1}}. It is seen that
when $t<0$ the two lumps are located in the first and third quadrants whereas they move to
the second and fourth quadrants after the collision, which is confirmed by Fig.~{\ref{3lumpasy1}}.
By a close observation, we find the evolutions of profile  in  Figs.~\ref{3lumpasy} and ~\ref{3lumpasy1} are opposite approximately along time $t$. For example,   Fig.~\ref{3lumpasy}(a) for t=-200 is corresponding to Fig.~\ref{3lumpasy1}(b) for $t=200$.  But, comparing with Eqs.(\ref{3lumpasytraj1}) and (\ref{3lumpasytraj4}), one can find that for the same $|t|$, these peak coordinates in the two cases are
not totally uniform. In other words, the modulus of the third-order lump $u^{[3]}$ with parameters (\ref{para1}) is not equal to that one
of the third-order lump $\widetilde{u^{[3]}}$ with parameters (\ref{para2}) at the same $|t|$ (i.e., $|u^{[3]}(x,y,t)|\neq|\widetilde{u^{[3]}}(x,y,-t)|$).
The heights of three peaks can be computed by inserting the approximate coordinates into the expression of this solution,
hence we do not repeat this step here.


\begin{remark}
 These cases further demonstrate that the scattering process
  does not necessarily occur at normal angles for the multi-lump
  solutions
  of the DS II equation, a key finding of the present work.
\end{remark}

\section{Conclusion and Discussion}

In this paper,  we showed the asymptotic properties of the newly obtained family of the
higher-order lump solutions for the DS II equation  in the moving coordinate frame (\ref{moving1}). For the higher-order
lump, when $|t|\rightarrow\infty$ we find that it splits into multi-peak lumps whose heights evolve
with time and approach the same constant value corresponding to that of the simple first-order
fundamental lump, and the peak trajectories have a time dependence
that grows as $\sqrt{|t|}$,  a feature
similar to what has been found for the KP I
equation\cite{Villarroel3,Villarroel4,PLA2000} and for higher-order
lumps on zero background of the DS II equation\cite{SIAM2003}.
Nevertheless, they define straight lines with different slopes for $t\rightarrow\pm\infty$.
The lumps are found to collide head-on undergoing a scattering process, and the
scattering angle $\Theta \in (\frac{\pi}{2}, \pi)$ is identified
herein as being different from the higher-order lump
on the zero background case where the scattering must be orthogonal. Besides, though we just discussed solutions up to the third-order,
$n$th-order lumps can be obtained by using the $n$-fold DT (\ref{un}).
Generalizing the results obtained herein to arbitrary $n$ would be an
interesting
topic for further study.

Our
results  concerning the dynamics for rational
solutions of DS II  can be a basis for corresponding
observations in areas of application where the DS II is relevant,
including
most notably in nonlinear optics and plasma physics, among others.
The method of construction and asymptotical analysis of exact
solutions of the
DS II in this paper can also
be widely used to other 2+1 dimensional integrable systems,
such as the KP equation, the $2+1$
 dimensional Fokas equation, etc.
Indeed, this prompts theoretical, numerical and even experimental
studies
to consider the angle of interaction of lump-like solutions that
can arise in settings that bear such solutions.

\begin{center}
{\bf Acknowledgments}
\end{center}
This work is supported by the National Natural Science Foundation of China (Grants 12071304 and 12101312).
the Natural Science Foundation of Jiangsu Province of China (Grant BK20210606) and the Natural Science Foundation of the Jiangsu
Higher Education Institutions of China (Grant 21KJB110030).
This
material is also based upon work supported by the US National
Science Foundation under Grant No. DMS-2204702 (P.G.K.).

\end{document}